\documentclass[12pt]{article}
\usepackage{graphicx}
\begin{document} 

\title{Long-range interactions in Sznajd consensus model}

\author{Christian Schulze\\
Institute for Theoretical Physics, Cologne University\\D-50923 K\"oln, Euroland}

\maketitle
\centerline{e-mail: ab127@uni-koeln.de}

\bigskip
Abstract: The traditional Sznajd model, as well as its Ochrombel 
simplification, for opinion spreading are modified to have a
convincing strength proportional to a negative power of the 
spatial distance. We find the usual phase transition in the 
full Sznajd model, but not in the Ochrombel simplification.
We also mix the two rules, which favours a phase transition.

\bigskip

Keywords: Sociophysics, phase transition, distance dependence,
quenched disorder 

PACS: 05.50 +q, 89.65 -s

\bigskip
Ising models have been studied since nearly one century, and
simulated on computers since more than four decades. A new 
version is the Sznajd model \cite{sznajd} where again each
lattice site carries a spin $\pm 1$. If two randomly selected
neighbouring spins have the same value, they force their 
neighbours to accept this value; otherwise nothing is 
changed and a new pair is selected. In the Ochrombel
simplification, instead of a pair, already a single site 
``convinces '' its neighbours \cite{ochrombel}. This model 
can be interpreted as the spreading of opinions until a consensus
is reached. Instead of two values $\pm 1$ we also can work with
$q$ values: 1,2, ..., $q$.  

The Sznajd model on the square lattice shows a phase transition:
If initially one of the two opinions has a slight majority in
a random distribution, then at the end all spins have that value
and the dynamics stops. The Ochrombel modification lost this 
transition \cite{schulze}. We now check for this transition in
the case of long-range interactions, decaying with a power law of
the distance, and with a mixture of Sznajd and Ochrombel rule.
The program is similar to the published one \cite{stauffer}.

 \begin{figure}[hbt]
\begin{center}
\includegraphics[angle=-90,scale=0.5]{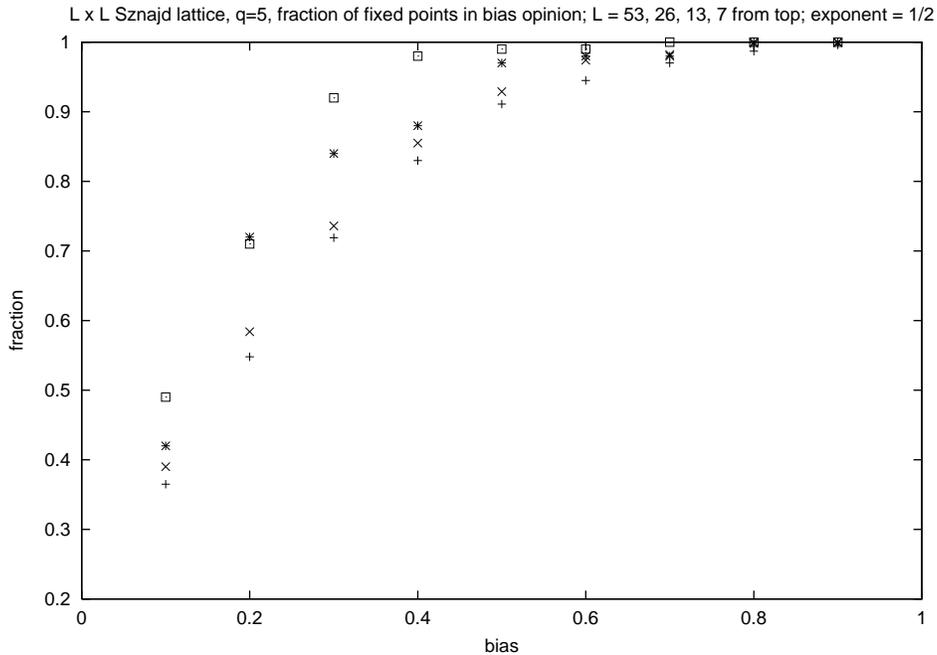}
\end{center}
\caption{Fraction of successes, $L=7$ to 53, exponent $x = 1/2$,
Sznajd case, five opinions}
\end{figure}

\begin{figure}[hbt]
\begin{center}
\includegraphics[angle=-90,scale=0.5]{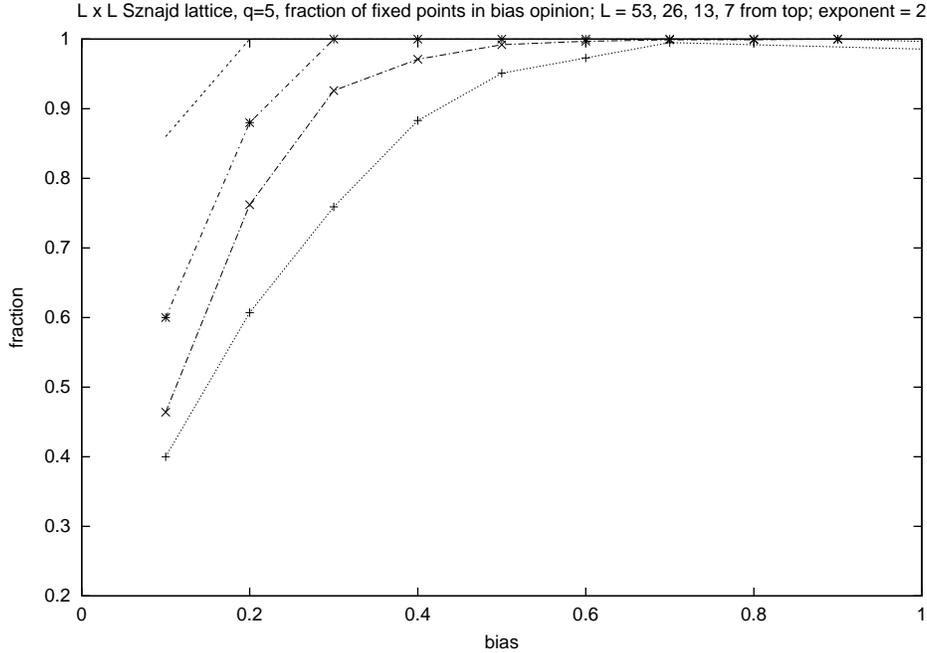}
\end{center}
\caption{As Fig.1 but with exponent $x = 2$ }
\end{figure}

We made 100 or 1000 simulations for $L \times L$ square lattices
with $L=7, 13, 26$ and 53, sometimes also 73,
usually allowing $q=5$ values. A spin
convinces, alone (Ochrombel) or together with an equally-minded 
neighbour (Sznajd), a neighbour at Euclidean distance $R$ with
probability $1/R^{2x}$. Initially the spins are distributed 
randomly among the $q$ opinions except that with a bias 
probability $p$ the just initialized spins are set to +1. A
quenched fraction $r$ of the sites follows the Sznajd pair rule,
the remaining fraction $1-r$ the Ochrombel single-site rule.
A success is a sample where at the end all spins had the bias
value +1.

Figs.1 and 2 show for the Sznajd case $r=1$
the phase transition: For
large $L$ a small bias $p$ suffices to make nearly all samples
successes. It does not matter much whether the interactions 
decay slowly ($x=1/2$) or fast ($x=2$) with distance. For the
$r=0$
Ochrombel case, however, analogous simulations (not shown) give
no phase transition, and this situation persists even if we 
take a very small $x = 0.1$ (Fig.3) and reduce $q$ from 5 to 2
(Fig.4).

Thus we mixed the two rules in Figs. 5 ($r=0.5$) and 6 ($r=0.1$)
which show a phase transition, for $x = 1/2$, in both 
cases. With a faster decay, $x=2$ instead of $x=1/2$, the phase
transition for $r=0.1$ becomes more pronounced, Fig.7.

In summary, contrary to our expectation from thermal phase 
transitions, the introduction of long-range interactions instead
of nearest-neighbour interactions did not create a phase
transition. 

\begin{figure}[hbt]
\begin{center}
\includegraphics[angle=-90,scale=0.5]{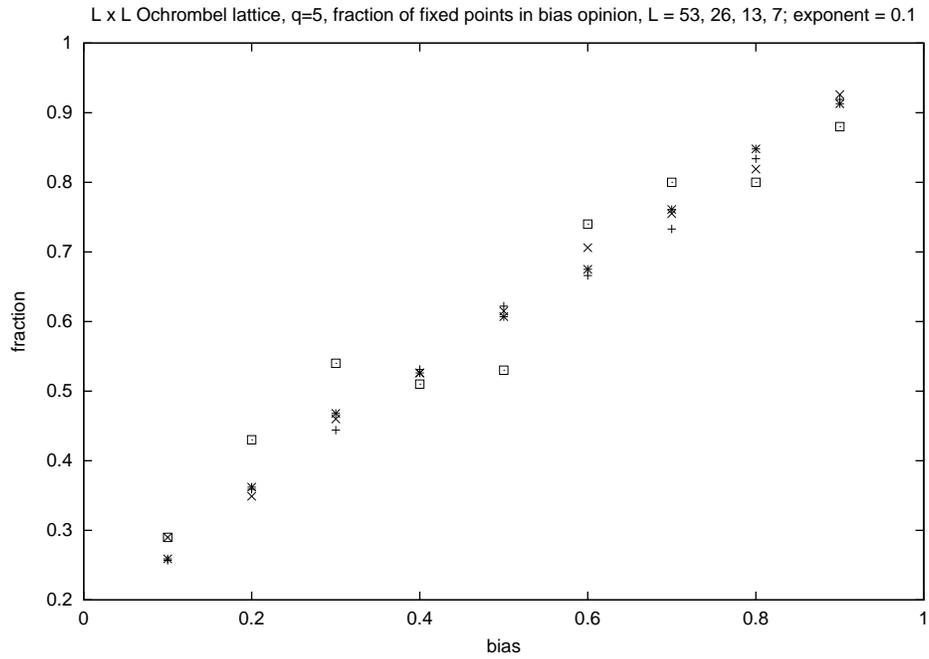}
\end{center}
\caption{ As Fig.1 but with exponent $x=0.1$, Ochrombel case}
\end{figure}

\begin{figure}[hbt]
\begin{center}
\includegraphics[angle=-90,scale=0.4]{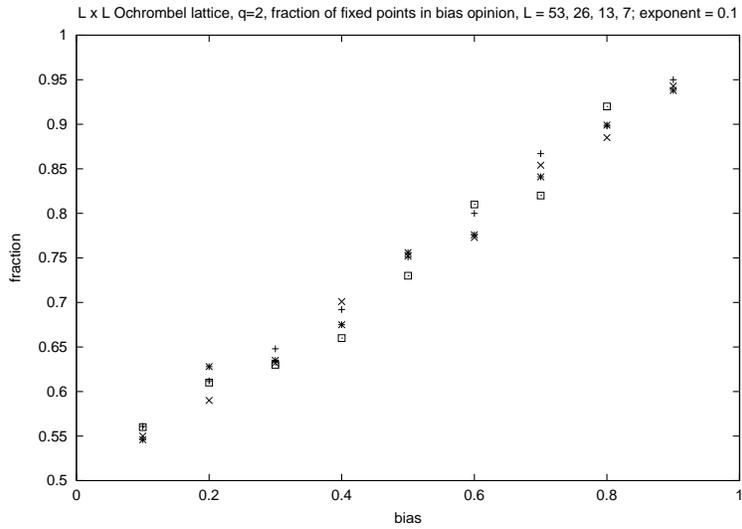}
\end{center}
\caption{ As Fig.3 but with exponent $x=0.1$, two opinions for
Ochrombel case}
\end{figure}

\begin{figure}[hbt]
\begin{center}
\includegraphics[angle=-90,scale=0.4]{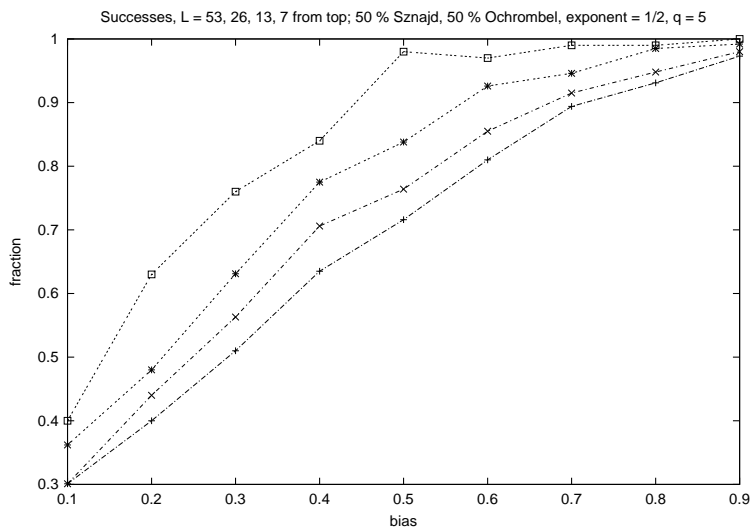}
\end{center}
\caption{ As Fig.1 (exponent $x=1/2$, five opinions) for
Sznajd-Ochrombel (50:50) mixture}
\end{figure}

\begin{figure}[hbt]
\begin{center}
\includegraphics[angle=-90,scale=0.4]{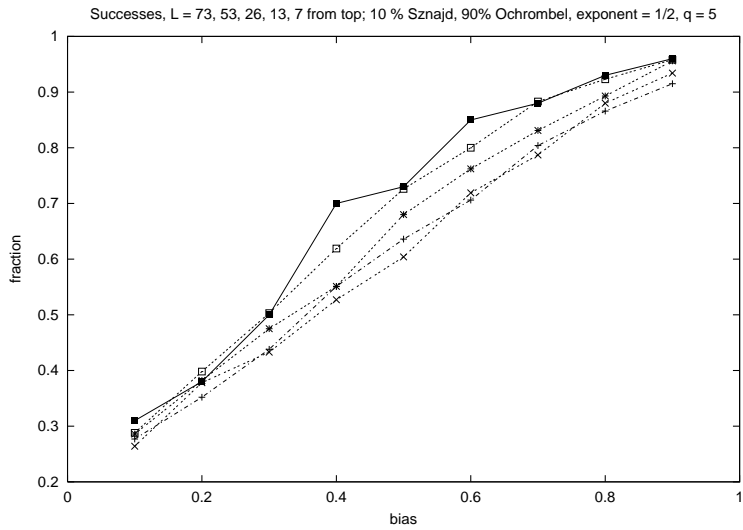}
\end{center}
\caption{ As Fig.5 (exponent $x=1/2$, five opinions) for
Sznajd-Ochrombel, but 10:90 mixture}
\end{figure}

\begin{figure}[hbt]
\begin{center}
\includegraphics[angle=-90,scale=0.4]{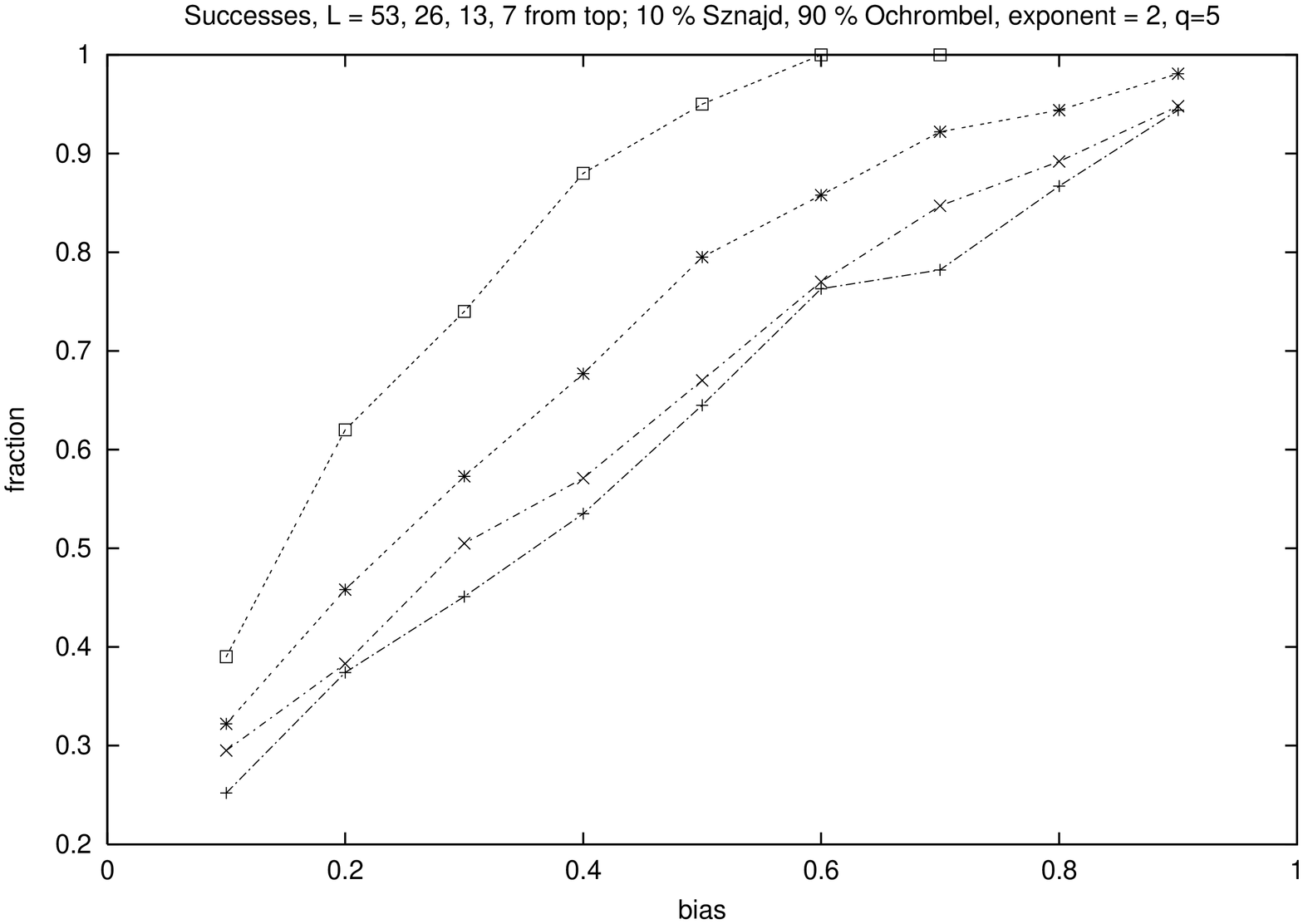}
\end{center}
\caption{ As Fig.6 (10:90 mixture, five opinions) for
exponent 2}   
\end{figure}

We thank Deutsche Forschungsgemeinschaft for support, 
and D. Stauffer, to whom this note is dedicated because of his
senility, for help.

\end{document}